\documentclass{article}

\usepackage{amsmath,amssymb,amsfonts,epsfig,psfrag,latexsym,lscape}
\usepackage{booktabs}
\usepackage[usenames,dvipsnames]{color}

\newcommand{\la}{\langle}
\newcommand{\ra}{\rangle}

\newcommand{\cT}{{\cal T}}
\newcommand{\cO}{{\cal O}}

\newcommand{\cL}{{\cal L}}

\newcommand\lr[1]{{\left({#1}\right)}}

\newcommand{\ft}[2]{{\textstyle\frac{#1}{#2}}}

\begin{document}

\thispagestyle{empty}

\begin{flushright}
HU-EP-16/25 
\end{flushright}

\begin{center}

\vskip 4.5 cm

{\Large \bf Half-BPS half-BPS twist two at four loops in ${\cal N} = 4$ SYM}

\vskip 2 cm
 
{\bf Burkhard Eden, Felix Paul} \\

\vskip 1 cm

{\it Institut f\"{u}r Mathematik und Physik, Humboldt-Universit{\"a}t zu Berlin, \\
Zum gro{\ss}en Windkanal 6, 12489 Berlin, Germany}

\vskip 1 cm

E-mail: eden@math.hu-berlin.de, paulfeli@physik.hu-berlin.de

\end{center}

\vskip 2.5 cm

We consider a double OPE limit of the planar four-point function of stress tensor multiplets in ${\cal N} = 4$ SYM theory. Loop integrands for this correlator have been constructed to very high order, but the corresponding integrals are explicitly known only up to three loops. Fortunately, the double coincidence limit of the four-loop integrals can be found by the method of expansion by regions, which reduces the problem of computing the four-point integrals to the evaluation of a large set of massless propagator integrals. These can in turn be evaluated by IBP reduction.

The OPE limit of the stress tensor four-point function allows us to extract the (square of the) three-point couplings between two stress tensor multiplets and one twist two operator in the $\textbf{20'}$ of $SU(4)$. The latest available IBP software accomplishes this task up to and including spin 8. With the data obtained we hope to further the development of the recent integrable systems picture for correlation functions.

\newpage

\section{Introduction}

The maximally supersymmetrically extended Yang-Mills theory --- ${\cal N} = 4$ SYM --- has many special properties: it is conformally invariant also on the quantum level, it is related to a certain string theory by the AdS/CFT correspondence, and the operator spectrum of the planar theory is described by an integrable system. Another recent discovery is an integrable system for the so-called remainder function in planar scattering amplitudes \cite{Alday:2009dv,Basso:2013vsa}. Indeed, one might conjecture that all other higher-point quantities of the planar theory can be captured by integrability, too.

The original object of interest within the context of the AdS/CFT duality were correlation functions of half-BPS operators, because their strong coupling asymptotics is provided by supergravity as a low energy limit of string theory. The four-point function of stress tensors is particularly well-studied. Naturally, one will ask whether such correlation functions --- or indeed $n$-point functions of more general gauge invariant composite operators --- can be found from integrability. Some ten years after the discovery of the integrability of the spectrum problem the ``hexagon proposal'' for computing three-point functions has been formulated in the recent breakthrough publication \cite{Basso:2015zoa}.

Three-point functions of half-BPS operators receive no quantum corrections \cite{D'Hoker:1998tz,Eden:1999gh}. Unfortunately, the direct computation of non-trivial three-point functions is highly non-trivial in perturbative field theory at higher loops, rendering it a hard task to vindicate any integrability prediction. On the other hand, OPE limits of BPS four-point functions make available certain classes of structure constants. In particular, the OPE coefficients between two half-BPS operators and one twist two operator have yielded precision tests at three loops \cite{Eden:2011we,Eden:2012rr,Drummond:2013nda,Chicherin:2015edu,Eden:2015ija,Basso:2015eqa}. In the present publication we return to the field theory side of this matching, but now at four loops.

The principal motivation for this study is a problem with the hexagon proposal \cite{Basso:2015zoa} that can be probed and hopefully mended using the structure constants that we elaborate here: the proposal contains the integration over the rapidity of virtual excitations. At four-loop level there can be two such virtual magnons. Their scattering leads to a double pole and thus a non-integrable singularity. Without field theory results to compare with it will be impossible to single out the right prescription to deal with this pole. 

\emph{Integrands} for the loop corrections to the planar four-point correlation function of stress tensors are known up to eight loops \cite{Eden:2011we,Eden:2012tu,Ambrosio:2013pba,Bourjaily:2015bpz}. However, the \emph{integrals} are unknown at four loops and above apart from a few exceptions. All in all, we encounter 26 genuine scalar conformal four-loop integrals. Some of these are related to the ladder graph by ``magic identities'' \cite{Drummond:2006rz}, others obey a Laplace equation \cite{Drummond:2012bg,Drummond:2013nda} or are linearly reducible \cite{Brown:2008um} and could therefore be evaluated by the publicly available code \emph{HyperInt} \cite{Panzer:2014caa}. But there is a core set whose analytic computation remains open. In \cite{Eden:2016dir} the differential equations method for master integrals of uniform transcendentality weight \cite{Kotikov:1990kg,Remiddi:1997ny,Henn:2013pwa} was applied in this context, for reasons of simplicity and comparability to begin with on a linearly reducible test case. The method is quite cumbersome, but it can, at least in principle, be used for all the integrals in the list.

On the other hand, the OPE coefficients of one twist two operator with two half-BPS multiplets can be determined by expansion by regions \cite{Smirnov:2002pj,Beneke:1997zp} with subsequent IBP reduction \cite{Chetyrkin:1981qh}. We employed the programs \emph{FIRE5} \cite{Smirnov:2014hma} and \emph{LiteRed} \cite{Lee:2012cn} for the reduction. The strategy was successful up to and including spin 8. The results of this paper confirm and extend the analysis of \cite{Goncalves:2016vir}. 

In Section \ref{sketch} we sketch the derivation of the loop integrands of the four-point function on the example of the three-loop contribution. Next, in Section \ref{ae} we discuss asymptotic expansions of scalar conformal four-point integrals. The technique of expansion by regions is  introduced for the one-loop box integral. We would like to refer the reader to \cite{Eden:2012rr} for more information, e.g. the asymptotic expansions of $E, H$ derived in this manner. Third, Section \ref{4l} displays the integrals of the four-loop contribution, although we cannot include much technical detail because the calculations were done by a \emph{Mathematica} script. Last, in  Section \ref{fin} we comment on the OPE decomposition of the four-point function and state the actual results.

\section{The stress tensor four-point function} \label{sketch}

Quantum corrections to the four-point function take a factorised form \cite{Eden:2000bk}:
\vskip - 0.4 cm
\begin{align}
  G_{4}(1,2,3,4)= G^{(0)}_{4}  + \frac{2 \, (N_c^2-1)}{(4\pi^2)^{4}} \  R(1,2,3,4)   \   \left[ a  F^{(1)} + a^2 F^{(2)} + a^3 F^{(3)} + a^4 F^{(4)} + O(a^5)  \right], \quad a = \frac{g^2 N}{16 \, \pi^2}.
\end{align}
Generically (as is the case in the tree contribution) the Clebsch-Gordan decomposition of the product of $SU(4)$ irreps carried by the stress tensors, i.e. $\mathbf{20'} \otimes \mathbf{20'} = \mathbf{1} \oplus \mathbf{15} \oplus \mathbf{20'} \oplus \mathbf{84} \oplus \mathbf{105} \oplus \mathbf{175}$, could lead to six distinct functions. The equation above says that these are indeed identical up to multiplication by rational factors. Explicitly,
\begin{align}
 R(1,2,3,4) &= \frac{y^2_{12}y^2_{23}y^2_{34}y^2_{14}}{x^2_{12}x^2_{23}x^2_{34} x^2_{14}}(x_{13}^2 x_{24}^2-x^2_{12} x^2_{34}-x^2_{14} x^2_{23}) +
\frac{ y^2_{12}y^2_{13}y^2_{24}y^2_{34}}{x^2_{12}x^2_{13}x^2_{24} x^2_{34}}(x^2_{14} x^2_{23}-x^2_{12} x^2_{34}-x_{13}^2 x_{24}^2)  \notag \\[2mm]
& + \, \frac{y^2_{13}y^2_{14}y^2_{23}y^2_{24}}{x^2_{13}x^2_{14}x^2_{23} x^2_{24}}(x^2_{12} x^2_{34}-x^2_{14} x^2_{23}-x_{13}^2 x_{24}^2) + 
\frac{y^4_{12} y^4_{34}}{x^2_{12}x^2_{34}}   + \frac{y^4_{13} y^4_{24}}{x^2_{13}
x^2_{24}} + \frac{y^4_{14} y^4_{23}}{x^2_{14}x^2_{23}} \label{rPoly}
\end{align}
with $x_{ij} = x_i - x_j$ and the square is a scalar product. Here the $\theta = \bar \theta = 0$ part of the correlator is written as on analytic superspace \cite{Galperin:1984av,Howe:1995md} which uses additional bosonic coordinates $y_{a a'}$ to keep track of the $SU(4)$ indices. 

The one- and two-loop integrands have been elaborated using Feynman graphs:
\begin{equation}
I_4^{(1)} \, \propto \, \frac{1}{x_{15}^2 x_{25}^2 x_{35}^2 x_{45}^2} \, , \qquad
I_4^{(2)} \, \propto \, \frac{x_{12}^2 x_{34}^2 x_{56}^2 \, + \, (14 \; \mathrm{terms})}
{(x_{15}^2 x_{25}^2 x_{35}^2 x_{45}^2) (x_{16}^2 x_{26}^2 x_{36}^2 x_{46}^2) x^2_{56}}
\end{equation}
Already at the two-loop level, the two existing calculations \cite{Eden:2000mv,Bianchi:2000hn} rely on symmetry considerations to simplify the otherwise very sizable task. The three-loop contribution would be exceedingly hard to obtain in a direct fashion. Fortunately, the form of $I_4^{(1)}, I_4^{(2)}$ suggests a pattern: The denominator has a factor $x_{1i}^2 x_{2i}^2 x_{3i}^2 x_{4i}^2$ for every integration vertex $x_i$ ($\{1,2,3,4\}$ label the outer points) and contains all links between integration points (above only $x_{56}^2$). At $l$ loops, the numerator is a polynomial of conformal weight\footnote{For weight $-n$ a point $i$ must occur exactly $n$ times in factors $x_{ij}^2$.} $- l+1$ at all points, and is --- somewhat surprisingly --- totally symmetric under the exchange of all points, regardless of whether they are external points or integration vertices \cite{Eden:2011we}. According to these principles an ansatz for the three-loop integrand is
\begin{equation}
  I_4^{(3)} \, \propto \, \frac{P^{(3)}}{(x_{15}^2 x_{25}^2 x_{35}^2 x_{45}^2)
(x_{16}^2 x_{26}^2 x_{36}^2 x_{46}^2) (x_{17}^2 x_{27}^2 x_{37}^2 x_{47}^2)
x^2_{56} x^2_{57} x^2_{67}} \nonumber
\end{equation}
where $P^{(3)}(x_{ij}^2)$ should be $S_7$ symmetric and it should have conformal weight -2 at every point. There are only four options:
\vskip - 0.5 cm
\begin{align}
 & {\text{(a) heptagon:}} &&x_{12}^2 x_{23}^2 x_{34}^2
  x_{45}^2 x_{56}^2 x_{67}^2  x_{71}^2  & \ +\ {S_7\ \mathrm{permutations}} \nonumber
 \\[1mm]
 & {\text{(b) 2-gon  $\times$ pentagon:}} &&(x_{12}^4)( x_{34}^2 x_{45}^2 x_{56}^2 x_{67}^2 x_{73}^2)  & \ +\ {S_7\ \mathrm{permutations}}  
 \nonumber
 \\[1mm]
   & {\text{(c) triangle $\times$  square:}} &&(x_{12}^2  x_{23}^2 x_{31}^2) ( x_{45}^2
  x_{56}^2 x_{67}^2 x_{74}^2 ) & \ +\ {S_7\ \mathrm{permutations}}   
  \nonumber
  \\[1mm]
 & {\text{(d) 2-gon $\times$ 2-gon $\times$ triangle:}} &&(x_{12}^4)( x_{34}^4)( 
   x_{56}^2 x_{67}^2 x_{75 }^2)  & \ +\ {S_7\ \mathrm{permutations}}
\nonumber
\end{align}
%
The three-loop correlator is thus fixed up to four constants. In fact, only polynomial (b) is allowed: this was originally seen \cite{Eden:2011we} by comparing to amplitude integrands via the correlator/amplitude duality \cite{Eden:2010zz}, from where also the coefficient of the polynomial can be taken over. Later on other criteria intrinsic to the correlator were developed \cite{Eden:2012tu}. Note that the correct normalisation of the polynomial is necessary for the exponentiation of the logarithms in a Euclidean OPE limit, see also Section \ref{fin} below. 

Remembering which points are outer points and integration vertices, respectively, we find the following functions in the quantum part up to three loops:
\begin{align}
F^{(1)} & =  g(1,2,3,4)\,,
\\[3mm]
 F^{(2)} & = h(1,2;3,4) + h(3,4;1,2) + h(2,3;1,4) +  h(1,4;2,3) 
 \\ & + \, h(1,3;2,4) + h(2,4;1,3)
  + \frac12 \lr{x_{12}^2x_{34}^2+ x_{13}^2 x_{24}^2+x_{14}^2x_{23}^2} 
[ \, g(1,2,3,4) ]\, ^2\, , \nonumber \\[3mm]
F^{(3)} & = \big[ \, L(1,3;2,4)+ 5 \mbox{ perms }\big] +
\big[ \, T(1,3;2,4)+ 11 \mbox{ perms }\big] \\[2mm] 
& + \, \big[ \, E(2;1,3;4)+ 11 \mbox{ perms }\big] +
{\textstyle \frac12}\big[ \, H(1,3;2,4)+ 11 \mbox{ perms }\big] \notag \\[2mm]
 & + \, \big[ \, ({g\times h})(1,3;2,4)+  5 \mbox{ perms }\big]
  \nonumber 
\end{align}
where the symbols mean the integrals
\begin{align}
& g(1,2,3,4)  = \frac{1}{2 \, \pi^2}
\int \frac{d^4x_5}{x_{15}^2 x_{25}^2 x_{35}^2 x_{45}^2}  \, , \label{box1}\\
& h(1,2;3,4)  =  \frac{x^2_{34}}{4 \, \pi^4}
\int \frac{d^4x_5 \, d^4x_6}{(x_{15}^2 x_{35}^2 x_{45}^2) x_{56}^2
(x_{26}^2 x_{36}^2 x_{46}^2)}
\end{align}
and at third order:
\begin{eqnarray}
{(g\times h)}(1,2;3,4) &=&{x_{12}^2 x_{34}^4\over 8 \, \pi^6}\int 
  \frac{d^4x_5d^4x_6 d^4x_7}{(x_{15}^2  x_{25}^2  x_{35}^2 x_{45}^2) 
 (x_{16}^2  x_{36}^2x_{46}^2) (x_{27}^2  x_{37}^2   x_{47}^2) x_{67}^2} \ , \\
L(1,2;3,4) & = &  \frac{x^4_{34}}{8 \, \pi^6}
\int \frac{d^4x_5 \, d^4x_6 \, d^4x_7}{(x_{15}^2 x_{35}^2 x_{45}^2) x_{56}^2
(x_{36}^2 x_{46}^2) x^2_{67} (x_{27}^2 x_{37}^2 x_{47}^2)} \, , \\
T(1,2;3,4)&=&{x_{34}^2\over 8 \, \pi^6}\int 
  \frac{ d^4x_5 d^4x_6 d^4x_7 \ x_{17}^2 }{(x_{15}^2 x_{35}^2)
(x_{16}^2  x_{46}^2) (x_{37}^2 x_{27}^2  x_{47}^2) x_{56}^2 x_{57}^2 x_{67}^2}\ , \\
E(1;3,4;2) & = &  \frac{x^2_{23} x^2_{24}}{8 \, \pi^6}
\int \frac{d^4x_5 \, d^4x_6 \, d^4x_7 \ x^2_{16}}{(x_{15}^2 x_{25}^2 x_{35}^2)
x_{56}^2 (x_{26}^2 x_{36}^2 x^2_{46}) x^2_{67} (x_{17}^2 x_{27}^2 x_{47}^2)}
\, , \label{eInt} \\
H(1,2;3,4) & = &  \frac{x_{34}^2 }{8 \, \pi^6}
\int \frac{d^4x_5 \, d^4x_6 \, d^4x_7 \ x^2_{57}}{(x_{15}^2 x_{25}^2 x_{35}^2
x^2_{45}) x_{56}^2 (x_{36}^2 x^2_{46}) x^2_{67} (x_{17}^2 x_{27}^2 x^2_{37}
x_{47}^2)}
\end{eqnarray}
In this list $g,h,L$ are the one-, two- and three-loop ladder graphs (or box integrals, referring to the planar dual), $T$ is the tennis court which is equal to $L$ in general kinematics as can be shown by ``flipping'' a sub-integral \cite{Drummond:2006rz}, and $E,H$ are new integrals that have been evaluated in terms of Goncharov polylogarithms only recently in \cite{Drummond:2013nda}.

\section{Asymptotic expansions of scalar conformal four-point integrals} \label{ae}

The entire series of $l$-loop box integrals is explicitly known \cite{Usyukina:1993ch}: one has
\begin{equation}
g(1,4,2,3) = \frac{1}{x_{13}^2 x_{24}^2} \Phi^{(1)}(u,v) \, , \qquad h(1,4;2,3) = \frac{1}{x_{13}^2 x_{24}^2} \Phi^{(2)}(u,v) \, , \qquad L(1,4;2,3) = \frac{1}{x_{13}^2 x_{24}^2} \Phi^{(3)}(u,v)
\end{equation}
etc. where
\vskip - 0.2 cm
\begin{equation}
u \, = \, \frac{x_{12}^2 x_{34}^2}{x_{13}^2 x_{24}^2} \, = \, x \bar x \, , \qquad
v \, = \, \frac{x_{14}^2 x_{23}^2}{x_{13}^2 x_{24}^2} \, = \, (1-x) (1-\bar x) \, ,
\end{equation}
\begin{equation}
\Phi^{(n)}(u,v) \, = \, \frac{(-1)^n}{x - \bar x} \, \sum_{r=0}^n \frac{(-1)^r (2 n - r)!}{r! (n-r)! n!}
\log^r(u) \left( \mathrm{Li}_{2 n - r}(x) - \mathrm{Li}_{2 n - r}(\bar x) \right) \, . \label{davy}
\end{equation}
The parametrisation of the conformal cross ratios $u,v$ by $x, \bar x$ makes this result particularly concise. In Euclidean kinematics $x, \bar x$ are complex and conjugate to each other.\footnote{In Minkowski kinematics they would be real but independent.} The ladder functions $\Phi^{(n)}$ have the very special property of being single-valued: the combination of logarithms and polylogarithms is such that all cuts cancel, note e.g. the combination $\log(x \bar x) = \log(u)$ with real and positive argument of the logarithm. They are the original example of single-valued harmonic polylogarithms (SVHPLs) \cite{BrownSVHPLs}. We can use the shuffle product for Goncharov logarithms to rewrite the ladder functions as sums over terms of the type $G_{w_1}(x) \, G_{w_2}(\bar x)$ where the weight vectors $w_i$ are formed from the two letters $0,1$. Single-valuedness implies that the entire sum can be reconstructed from the part with no $\bar x$ dependence by requiring the absence of discontinuities. Labeling every SVHPL by the weight vector of the characteristic $G_w(x)$ term one finds
\begin{equation}
\Phi^{(1)}(u,v) = \frac{\cL_{1,0}-\cL_{0,1}}{x-\bar x} \, , \qquad 
\Phi^{(2)}(u,v) = \frac{\cL_{0,1,0,0}-\cL_{0,0,1,0}}{x-\bar x} \, , \qquad
\Phi^{(3)}(u,v) = \frac{\cL_{0,0,1,0,0,0}-\cL_{0,0,0,1,0,0}}{x-\bar x} \label{ladderSVHPLs}
\end{equation}
from which it is easy to infer how the series goes on. The $E$ integral can be concisely written in terms of SVHPLs, too, while $H$ is also single-valued but has a more complicated alphabet for the weight vectors \cite{Drummond:2013nda}.

In order to extract structure constants we will employ a double Euclidean limit $x_{12}, x_{34} \to 0$. Clearly,
$u \to 0, \, v \to 1$ in this limit, so one considers a double expansion in the small parameters $u$ and $Y = 1 - v$. Using the parametrisation $x,\bar x$ the limit can simply be realised by taking $\bar x$ to zero. The ladder integrals exemplify that one obtains an expansion in $\log(u)$ and $Y = x$ in this way. Here the maximum power of $\log(u)$ is equal to the loop order. In this letter we will not be interested in power corrections in $u$. On the other hand, $\log(Y)$ does not occur whereas we wish to derive a power series in $Y$. 

Conformal covariance implies that each of $g,h,L,T,E$ has the form
\begin{equation}
I_{1234} = \frac{1}{x_{13}^2 x_{24}^2} \, f(u,v) \label{confForm}
\end{equation}
if the outer factors $x_{ij}^2$ in front of the integral signs in (\ref{box1}) to (\ref{eInt}) are included into the definition. ($H$ as defined above is of conformal weight 2 at each point.) We can remove, say, point 4 by the limit
\begin{equation}
\lim_{x_4 \to \infty} x_4^2  \, I_{1234} = \frac{1}{x_{13}^2} \, f(u',v') \, , \qquad u' = \frac{x_{12}^2}{x_{13}^2}, \, v' = \frac{x_{23}^2}{x_{13}^2} \, .
\end{equation}
which does not loose information, because $x_{13}^2$ can uniquely be completed to $x_{13}^2 x_{24}^2$ and so on. On propagator representations as in (\ref{box1}) to (\ref{eInt}) the limit means to cancel every line (in denominator and numerator) connected to point 4. Because of the possibility to reconstruct it is unimportant which outer point is amputated in this way; generally one will choose the one with most lines attached. The result is a scalar three-point integral with fewer propagators.

There are only six distinct point permutations of (\ref{confForm}) under the $S_4$ group permuting $\{1,2,3,4\}$. The limit breaks the full $S_4$ to the $S_3$ permutation group of the remaining three points which still produces, of course, all six possible permutations of (\ref{confForm}). If the integral in question has some extra symmetry beyond conformal invariance then some of the six asymptotic expansions will be equal.

For integrals which are not explicitly known the Euclidean coincidence limit can be obtained by ``expansion by regions'' \cite{Smirnov:2002pj,Beneke:1997zp}. We will sketch the method on the example of the one-loop box. To this end, we send point 4 to infinity, put point 1 to zero and put $p_1 := x_2, \, p_2 := x_3$. Now, let $|p_1| << |p_2|$ whereby this is a Euclidean limit $x_{12} \to 0$.  We find
\begin{equation}
u' = \frac{p_1^2}{p_2^2} \, , \qquad v' = 1 - \frac{2 \, p_1 . p_2}{p_2^2} + u' \, , \qquad g(1,2,3,4) \, \to \, I' = \frac{1}{2 \pi^2} \int \frac{d^Dk}{k^2 (k-p_1)^2 (k-p_2)^2}
\end{equation}
and hence the integral becomes a massless vertex graph, and to leading order in $u'$ we can identify $Y = 2 \, p_1 . p_2 / p_2^2$.

In the ``hard'' or ``top region'' we assume $k \sim O(p_2)$ and expand
\begin{equation}
\frac{1}{(k-p_1)^2} \, = \, \frac{1}{k^2} \sum_{n=0}^\infty \left( \frac{2 \, k . p_1}{k^2} \right)^n + O(p_1)^2 \, .
\end{equation}
Extending the integration domain all the way to $|k| = 0$ is of course not compatible with this expansion; every term in the series is divergent at the origin. Instead of modifying the expansion as one would e.g. in the Gegenbauer technique we rather keep it and dimensionally regularise the infrared divergences:
\begin{equation}
I' \, \rightarrow \, \frac{1}{2 \pi^2} \int \frac{d^Dk}{(k^2)^2 (k-p_2)^2} \sum_{n=0}^\infty \left( \frac{2 \, k.p_1}{k^2} \right)^n + O(p_1^2) \, = \, \frac{1}{(p_2^2)^{1 + \epsilon}} \, \sum_{n=0}^\infty G(2 + n,1,n) \ Y^n + O(u) \label{hardBox1}
\end{equation}
The remaining one-loop bubble integral (not a triangle anymore!) can be exactly evaluated
\begin{equation}
\frac{1}{2 \pi^2} \int \frac{d^Dk \ k^{\mu_1} \ldots k^{\mu_n}}{(k^2)^\alpha \, ((k-q)^2)^\beta} \, = \, G(\alpha, \beta, n) \, \frac{q^{\mu_1} \ldots q^{\mu_n}}{(q^2)^{\alpha + \beta - D/2}} \, + \, \mathrm{traces} \, , \qquad D \, = \, 4 - 2 \, \epsilon \, , \label{explicitBubble}
\end{equation}
\begin{equation}
G(\alpha,\beta,n) \, = \,
\frac{\Gamma(\alpha + \beta - D/2) \Gamma(D/2 - \alpha + n) \Gamma(D/2 - \beta)}{\Gamma(\alpha) \Gamma(\beta) \Gamma(D - \alpha - \beta + n)}
\end{equation}
(c.f. \cite{mincer} and references therein) which yields the r.h.s. of equation (\ref{hardBox1}).

Similarly, the ``soft'' or ``bottom region'' $k \sim O(p_1)$ is done expanding
\begin{equation}
\frac{1}{(k-p_2)^2} \, = \, \frac{1}{p_2^2} \sum_{n=0}^\infty \left( \frac{2 \, k . p_2}{p_2^2} \right)^n + O(k^2) \, .
\end{equation}
\begin{equation}
I' \, \rightarrow \, \frac{1}{2 \pi^2 \, p_2^2} \, \int \frac{d^Dk}{k^2 (k-p_1)^2} \sum_{n=0}^\infty \left( \frac{2 \, k.p_2}{p_2^2} \right)^n + O(u) \, = \, \frac{1}{p_2^2 \, (p_1^2)^\epsilon} \, \sum_{n=0}^\infty G(1,1,n) \ Y^n + O(u)
\end{equation}
where the dimensional regulator mends the a priori incorrect extension of the integration domain to $|k| \to \infty$, so the ultraviolet divergences of the individual terms of the series.

Now,
\begin{equation}
G(1,1,n)|_{\epsilon^{-1}} \, = \, \frac{1}{(n+1) \, \epsilon} \, = \, - G(2+n,1,n)|_{\epsilon^{-1}} \, , \quad
G(1,1,n) \, + \, G(2+n,1,n) \, = \, \frac{2}{(n+1)^2} \, + \, O(\epsilon)
\end{equation}
so that
\begin{equation}
\mathbf{\mathrm{Hard}} + \mathbf{\mathrm{Soft}} \, = \, \, \frac{1}{p_2^2} \, \sum_{n=1}^\infty \frac{Y^{n-1}}{n} \left[ - \log(u) + \frac{2}{n} \right] + O(u) \, . \label{hardSoft}
\end{equation}
Since the one-loop box is finite in general kinematics we only need the leading $O(\epsilon^0)$ term. We put the regulator to zero once the two regions have been added. The result (\ref{hardSoft}) agrees with the asymptotic expansion of $\Phi^{(1)}$ by the $\bar x = 0$ trick. For our vertex integrals, the method yields asymptotic expansions (power corrections in $u$ could be included on the expense of working harder) but never a closed form solution like (\ref{ladderSVHPLs}). On the other hand, this is exactly what we will need to extract structure constants.

For $l$-loop vertex integrals we distinguish $2^l$ regions according to whether $k_i \sim O(p_1)$ or $k_i \sim O(p_2)$. In the hard region (all loop momenta assumed $\sim O(p_2)$) we find $l$-loop bubble integrals with high indices\footnote{exponents of denominator factors} for some of the lines. Similarly, the soft region (all loop momenta assumed $\sim O(p_1)$) yields $l$-loop bubble integrals, although with all indices smaller or equal 1. Massless bubble integrals are an ideal playground for the application of IBP techniques \cite{Chetyrkin:1981qh}; up to and including three loops there is the \emph{Mincer} system \cite{mincer}, for more general/higher loop tasks we used \emph{FIRE5} \cite{Smirnov:2014hma} in conjunction with \emph{LiteRed} \cite{Lee:2012cn}. Due to the high indices the hard region is also the hardest one to solve, even the strongest IBP systems come to a halt at still moderately high orders in $Y$.

In mixed regions the vertex integral breaks into one bubble integral with denominators depending on $p_2$ and another one depending on $p_1$. However, the geometric series generate numerators with dot products between large and small momenta (also for the hard and soft region). By tensor reduction these can be re-expressed by powers of $Y = 2 p_1.p_2/p_2^2$ times scalar bubble integrals of only $p_1$ or $p_2$ respectively, which can then be fed into the aforementioned IBP systems. In case of the one-loop box this problem did not arise only because formula (\ref{explicitBubble}) can handle integrals with open indices. However, there is no such result for more complicated topologies.

A simple algorithm for the tensor reduction at leading order in $u$ and a much more complete description of the method of expansion by regions as applied to this class of integrals is given in \cite{Eden:2012rr}, where the asymptotic expansions of $E$ and $H$ at lowest order in $u$ were obtained in closed form in terms of harmonic sums. This was possible because the Mincer system could solve the three-loop hard regions up to $Y^{33}$ which was sufficient to obtain a fit. In the four-loop case at hand we are not so lucky: \emph{FIRE5} with \emph{LiteRed} was at ease up to $Y^6$ while a preliminary run\footnote{We are grateful to V.~Smirnov for this test.} suggested problems already at $Y^8$.

Last, the limits (i) $x_1 = 0, x_2 = p_1, x_3 = p_2$ and (ii) $x_1 = p_1, x_2 = 0, x_3 = p_2$ with $|p_1| << |p_2|$ both imply $x_{12} \to 0$. Nevertheless, in general these are inequivalent: for an integral of conformal weight 1 at all points limit (ii) is obtained from (i) by the variable transformation $u \mapsto u/(1-Y), \, Y \mapsto Y/(Y-1)$ and subsequent division of the entire series by $1-Y$. We have used this as a consistency check for all asymptotic expansions at four loops; the six limits of every integral are indeed pairwise connected by this transformation.

\section{Four loops} \label{4l}

The planar part of the four-point integrand is given by the three polynomials \cite{Eden:2012tu}
\begin{align} \notag
P^{(4)}_1(x_1, \dots, x_8)&=\ft1{24} x_{12}^2 x_{13}^2 x_{16}^2 x_{23}^2 x_{25}^
2 x_{34}^2 x_{45}^2 x_{46}^2 x_{56}^2 x_{78}^6 & + \ \text{$S_8$ permutations}\,,  \\ 
P^{(4)}_2(x_1, \dots, x_8)&=\ft18x_{12}^2 x_{13}^2 x_{16}^2
   x_{24}^2 x_{27}^2 x_{34}^2 x_{38}^2 x_{45}^2 x_{56}^4 x_{78}^4 & + \ \text{$S_8$
   permutations}\,,\label{fourPolys} \\
P^{(4)}_3(x_1, \dots, x_8)&=\ft1{16} x_{12}^2 x_{15}^2 x_{18}^2 x_{23}^2 x_{26}^
2 x_{34}^2
   x_{37}^2 x_{45}^2 x_{48}^2 x_{56}^2 x_{67}^2 x_{78}^2 & + \ \text{$S_8$ permutations} \notag
 \end{align}
with coefficients $\{1,1,-1\}$, respectively. The weight of the symmetrisation in (\ref{fourPolys}) is such that every inequivalent term occurs only once. Taking out a factor $4!$ for the permutation symmetry of the integration vertices we find
\begin{align}
 F^{(4)} ={1\over 4!\,(2 \pi^2)^4} \int \frac{d^4x_5   d^4x_6 d^4
x_7d^4x_8\, \left[ P^{(4)}_1(x_i)+P^{(4)}_2(x_i)-P^{(4)}_3(x_i)\right]}{x_{56}^2
x_{57}^2x_{58}^2x_{67}^2x_{68}^2x_{78}^2\prod_{i=1}^4 x_{i5}^2 x_{i6}^2x_{i7}^2x
_{i8}^2}\,. \label{f4}
\end{align}
Keeping track of which points are outer points and integration vertices, respectively, one obtains a list of six disconnected and 26 connected --- so genuinely four-loop --- integrals. As is required by the planar correlator/amplitude duality (which states roughly that the correlator integrand generates the square of amplitude integrands), the disconnected part only shows one-loop times three-loop and two-loop times two-loop integrals.  These are
\begin{align}
D_1 & =  x^2_{14} x^2_{23} \; g(1,2,3,4) \, s(1,2;3,4) \, , \nonumber \\
D_2 & =  x^2_{12} x^2_{34} \; g(1,2,3,4) \, T(1,2;3,4) \, , \nonumber \\
D_3 & =  x^2_{12} x^2_{34} \; g(1,2,3,4) \, L(1,2;3,4) \, , \\
D_4 & =  x^2_{12} x^2_{34} \; h(1,2;3,4) \, h(1,2;3,4) \, , \nonumber \\
D_5 & =  x^2_{14} x^2_{23} \; h(1,2;3,4) \, h(3,4;1,2) \, , \nonumber \\
D_6 & =  x^2_{12} x^2_{34} \; h(1,2;3,4) \, h(3,4;1,2) \, .
\end{align}
Now, $h(1,2;3,4) = h(3,4;1,2)$ due to conformal invariance as in (\ref{confForm}), which is the ``flip identity'' that implies the equality of three-loop ladder and tennis court \cite{Drummond:2006rz} when applied to a subintegral. Hence $D_2 = D_3$ and $D_4 = D_6$. The integral $s$ is a new three-loop case:
\begin{equation}
s(1,2;3,4) = \frac{x_{34}^2}{8 \, \pi^6} \int \frac{d^4x_5 d^4 x_6 d^4x_7}{(x_{15}^2 x_{35}^2) x_{56}^2 (x_{26}^2 x_{46}^2) (x_{37}^2 x_{47}^2) x_{57}^2 x_{67}^2}
\end{equation}
As points 1,2 are connected to the integration vertices by only one line it can be solved by the Laplace technique of \cite{Drummond:2012bg}. Like for the ladder integrals a concise expression in terms of SVHPLs is found \cite{stephan}:
\begin{equation}
s(1,2;3,4) = \frac{-\cL_{1, 0, 0, 1, 0, 1} + \cL_{1, 0, 1, 0, 0, 1} + 
 4 \zeta_3 \, \cL_{1, 0, 1} - 20 \zeta_5 \, \cL_{1}}{x_{13}^2 x_{24}^2 \, (x - \bar x)}
\end{equation}
Note that this integral has six distinct Euclidean coincidence limits. This also happens in case of $D_5$ due to the ``oblique'' outer rational factor. To assemble the contributions of the $D_i$ we count the number of integrals of each type, divide by the factor of $4!$ in (\ref{f4}) and then another factor of 6. Including the signs of $P^{(4)}_j$ this yields the vector of coefficients $\{-2, 2, 1, 1/2, -1, 1/2\}$ for the sum of all six limits of each $D_i$.

The genuine four-loop integrals are
\begin{align}
N_1 &= x^6_{18} x^2_{25} x^2_{26} x^2_{27} x^2_{34} x^2_{36} x^2_{37} x^2_{45} x^2_{47} x^2_{56} \, , \notag \\ 
N_2 &= x^2_{14} x^4_{18} x^2_{25} x^2_{27} x^2_{28} x^4_{36} x^2_{37} x^2_{45} x^2_{47} x^2_{56} \, , \notag \\ 
N_3 &= x^2_{15} x^4_{18} x^2_{24} x^2_{26} x^2_{28} x^2_{36} x^4_{37} x^2_{45} x^2_{47} x^2_{56} \, , \notag \\ 
N_4 &= x^2_{16} x^2_{17} x^2_{18} x^2_{25} x^2_{27} x^2_{28} x^2_{34} x^2_{36} x^2_{38} x^2_{45} x^2_{47} x^2_{56} \, , \notag \\ 
N_5 &= x^2_{16} x^2_{17} x^2_{18} x^2_{25} x^2_{27} x^2_{28} x^4_{34} x^2_{38} x^2_{47} x^4_{56} \, , \notag \\ 
N_6 &= x^4_{17} x^2_{18} x^2_{24} x^2_{26} x^2_{28} x^2_{34} x^2_{35} x^2_{38} x^2_{47} x^4_{56} \, , \notag \\ 
N_7 &= x^2_{14} x^2_{17} x^2_{18} x^2_{23} x^2_{27} x^2_{28} x^2_{34} x^2_{38} x^2_{47} x^6_{56} \, , \notag \\ 
N_8 &= x^2_{15} x^4_{18} x^2_{26} x^2_{27} x^2_{28} x^4_{34} x^2_{37} x^2_{46} x^2_{56} x^2_{57} \, , \notag \\ 
N_9 &= x^6_{18} x^2_{24} x^2_{26} x^2_{27} x^2_{34} x^2_{35} x^2_{37} x^2_{46} x^2_{56} x^2_{57} \, , \notag \\ 
N_{10} &= x^2_{16} x^4_{18} x^2_{24} x^2_{25} x^2_{28} x^2_{34} x^4_{37} x^2_{46} x^2_{56} x^2_{57} \, , 
\notag \\
N_{11} &= x^2_{16} x^2_{17} x^2_{18} x^2_{24} x^2_{27} x^2_{28} x^2_{34} x^2_{35} x^2_{38} x^2_{46} x^2_{56} x^2_{57} \, , \notag \\
N_{12} &= x^2_{13} x^4_{18} x^2_{24} x^2_{27} x^2_{28} x^2_{34} x^2_{37} x^2_{46} x^4_{56} x^2_{57} \, , \notag \\ 
N_{13} &= x^6_{18} x^2_{23} x^2_{24} x^2_{27} x^2_{34} x^2_{36} x^2_{45} x^2_{56} x^2_{57} x^2_{67} \, , 
\end{align}
\begin{align}
N_{14} &= x^2_{14} x^2_{17} x^2_{18} x^2_{23} x^2_{26} x^2_{28} x^2_{35} x^2_{37} x^2_{45} x^2_{46} x^2_{58} x^2_{67} \, , \notag \\ 
N_{15} &= x^2_{16} x^2_{17} x^2_{18} x^2_{25} x^2_{27} x^2_{28} x^6_{34} x^2_{56} x^2_{58} x^2_{67} \, , \notag \\ 
N_{16} &= x^4_{17} x^2_{18} x^2_{24} x^2_{26} x^2_{28} x^4_{34} x^2_{35} x^2_{56} x^2_{58} x^2_{67} \, , \notag \\ 
N_{17} &= x^2_{14} x^2_{17} x^2_{18} x^2_{24} x^2_{26} x^2_{28} x^2_{34} x^2_{35} x^2_{37} x^2_{56} x^2_{58} x^2_{67} \, , \notag \\ 
N_{18} &= x^2_{13} x^4_{18} x^2_{24} x^4_{27} x^2_{34} x^2_{36} x^2_{45} x^2_{56} x^2_{58} x^2_{67} \, , \notag \\
N_{19} &= x^2_{14} x^2_{17} x^2_{18} x^2_{23} x^2_{27} x^2_{28} x^2_{34} x^2_{36} x^2_{45} x^2_{56} x^2_{58} x^2_{67} \, , \notag \\ 
N_{20} &= x^2_{14} x^2_{17} x^2_{18} x^2_{23} x^2_{26} x^2_{28} x^2_{34} x^2_{37} x^2_{45} x^2_{56} x^2_{58} x^2_{67} \, , \notag \\ 
N_{21} &= x^2_{14} x^2_{17} x^2_{18} x^2_{23} x^2_{27} x^2_{28} x^4_{34} x^4_{56} x^2_{58} x^2_{67} \, , \notag \\ 
N_{22} &= x^2_{14} x^2_{16} x^2_{18} x^2_{24} x^4_{27} x^2_{34} x^2_{35} x^2_{36} x^4_{58} x^2_{67} \, , \notag \\ 
N_{23} &= x^2_{13} x^2_{17} x^2_{18} x^4_{24} x^2_{27} x^2_{34} x^2_{36} x^2_{56} x^4_{58} x^2_{67} \, , \notag \\ 
N_{24} &= x^2_{14} x^2_{16} x^2_{17} x^2_{23} x^2_{24} x^2_{27} x^2_{34} x^2_{36} x^6_{58} x^2_{67} \, , \notag \\ 
N_{25} &= x^2_{12} x^2_{14} x^2_{17} x^2_{23} x^2_{27} x^2_{34} x^2_{36} x^2_{46} x^6_{58} x^2_{67} \, , \notag \\ 
N_{26} &= x^2_{12} x^2_{13} x^2_{18} x^2_{24} x^2_{27} x^2_{34} x^2_{36} x^2_{45} x^4_{58} x^4_{67} \, . \notag
\end{align}
with
\begin{equation}
I_j(1,2,3,4) = \frac{1}{16 \pi^8} \int \frac{d^4x_5 d^4x_6 d^4x_7 d^4x_8 \; N_j}{x_{56}^2
x_{57}^2x_{58}^2x_{67}^2x_{68}^2x_{78}^2\prod_{i=1}^4 x_{i5}^2 x_{i6}^2x_{i7}^2x_{i8}^2}
\end{equation}
As before, we determine combinatorical coefficients by dividing the number of integrals of a given type by 144 and dragging along the signs. One finds the vector
\begin{equation}
  \{2, 4, 2, -2, 1, 2, 1, 2, 4, 4, -4, 4, 2/3, -(1/2), 1,
   4, -2, 2, -2, -4, 1, 2, 4, 1, 1/2, 1/2\}
\end{equation}
of weights for the symmetric sum of the six asymptotic expansions of each integral. Note that $I_1, I_2, I_5, I_8$ are identically equal to the four-loop ladder integral $I_{15}$ by flipping subintegrals \cite{Drummond:2006rz}. We may use (\ref{davy}) at $\bar x = 0$ in these cases, while the other integrals can only be addressed by expansion by regions.

We have used \emph{Mathematica} to derive the relevant set of scalar bubble integrals in all regions of the integrals and then evaluated two- and three-loop bubble integrals on a laptop using the \emph{Mathematica} version of \emph{FIRE5} with \emph{LiteRed} rules. \emph{LiteRed} can derive recursion rules for all massless three-loop bubble topologies. At four loops this is unfortunately not so. The top and bottom regions of the integrals $I_{14}, I_{16} \ldots I_{26}$ all fit into a single family of four-loop propagator bubbles with 3 out of 14 indices non-positive. \emph{LiteRed} is able to find reduction rules for this top sector of the problem. The IBP reduction of all $\approx$ 55000 scalar bubble integrals occurring in the asymptotic expansion of these integrals up to order $Y^6$ was attempted in a single run using the \emph{C++} version of \emph{FIRE5} with \emph{LiteRed} rules and eventually took some two days on an AMD blade with 48 kernels and 256 GB RAM. The top and bottom regions of the other integrals can be sorted into six further families of four-loop bubbles, though with four non-positive indices. \emph{LiteRed} finds complete sets of recursion rules also in these sectors. We have likewise collected all integrals needed up to $Y^6$ and run a single reduction for each family. There are less integrals in these sectors and the reduction is simpler because of the four non-positive indices.

The relevant master integrals are listed in \cite{Baikov:2010hf} (see also \cite{Lee:2011jt}) barring for two exceptions specific to configuration space. The latter also occurred in the evaluation of the planar five-loop anomalous dimension of the Konishi multiplet and have been worked out to the relevant order in the $\epsilon$ expansion in \cite{gamma5}.

\section{Waves and Conclusions} \label{fin}

In the double coincidence limit $x_{12}, x_{34} \to 0$ the four-point function at $\theta = \bar \theta = 0$ can be decomposed into an infinite sum over ``conformal partial waves'' (CPWA) correponding to the two OPEs $\la \cT_1 \cT_2 \cO \ra$ and $\la \cT_3 \cT_4 \cO \ra$, and a two-point function of the ``exchanged'' operator $\cO$ of dimension $\Delta = \Delta_0 + \gamma$ and spin $s$:
\begin{equation} 
G_4 \, \propto \, \sum_{\gamma,s} \, \left(N_{\cT \cT}^{\cO}\right)^2 \, {\text{cpwa}}(\gamma,s) \label{cpwa1}
\end{equation}
One has $\Delta_0 \in \mathbb{N}, \, \gamma = a \gamma_1 + a^2 \gamma_2 + a^3 \gamma_3 + a^4 \gamma_4 + \ldots$. The integer part $\Delta_0$ is the classical scaling weight of the exchanged operator, while the coupling dependent ``anomalous'' part has to be computed by quantum field theory. The ``twist'' of an operator is defined as $\Delta_0 - s \geq 2$. Twist two conformal blocks take the form \cite{pisa,glebOPE,osborn0} (and references therein)
\begin{equation}
\text{cpwa}(\gamma,s) \, = \, u^{\gamma/2} \; Y^s \; 
_2F_1 \Bigl(s+1+ \gamma/2, s + 1 + \gamma/2, 2 + 2 \, s + \gamma; \, Y \Bigr) \, . \label{cpwa2}
\end{equation}
Expanding the anomalous part of the dimension in the effective coupling constant $a$ creates logarithms of $u$ which can be matched on those in the asymptotic expansion of the Feynman integrals. The $\la \cT_1 \cT_2 \cO \ra$ OPE contains operators of any even twist; the corresponding CPWA come with extra powers of $u$. The restriction to lowest order in $u$ therefore selects the twist two trajectory.

Note that the $Y$ expansion of a spin $s$ CPWA starts on $Y^s$. Matching on the asymptotic expansion of the perturbative corrections to the four-point functions thus produces a triangular system of equations  which one will truncate at the desired order in $a, Y$. This can easily be solved by back-substitution. Second, there is only one new twist two operator at every even spin. Therefore we obtain a unique solution for every
\begin{equation}
\left(N_{\cT \cT}^{\cO_{2+s+\gamma}^s}\right)^2 = \alpha_0(s) + a \, \alpha_1(s) + a^2 \, \alpha_2(s) + a^3 \, \alpha_3(s) + a^4 \, \alpha_4(s) + \ldots \, , \qquad s \in 2 \mathbb{N}
\end{equation}	
while the structure constants vanish for odd spin. On the other hand, structure constants for twist four operators, say, could not be inferred from the single correlator $\la \cT \cT \cT \cT \ra$. This is our reason to restrict to leading twist.

Last, the $R$ polynomial of equation (\ref{rPoly}) as well as the tree-level correlator contain six $y$ structures corresponding to the six summands in the Clebsch-Gordan series of the product of $SU(4)$ irreps $\mathbf{20'} \otimes \mathbf{20'}$. Here we wish to focus on the exchange of twist two operators in the $\mathbf{20'}$ channel. The corresponding $SU(4)$ projector has been given in \cite{glebOPE}. Applied to the tree-level contribution and the $R$ polynomial, respectively, we find \cite{Eden:2012rr}
\begin{equation}
\lim_{x_{12}, x_{34} \rightarrow 0} 
G_4^\mathbf{20'} =  \frac{5 N^2}{3 \, (4 \pi^2)^4 \, x_{13}^2 x_{24}^2 \, u} \left( \frac{2-Y}{1-Y} -  2 \frac{Y^2}{1-Y} \sum_{l = 1}^\infty \lim_{x_{12}, x_{34}
\rightarrow 0} \, a^l \, F^{(l)}(x_i)  + O(u) \right) \label{corLim}
\end{equation}
in the planar limit. The part in the round brackets is to be equated with the r.h.s. of (\ref{cpwa1}). The numerator factor $Y^2$ in front of the $a^l F^{(l)}$ makes the interacting part given by the Feynman integrals contribute at higher orders than expected. As a result we can read off four-loop anomalous dimensions and structure constants up to spin 8.

For spin 0 there are no loop corrections to the dimension or structure constant. We find for the four-loop anomalous dimensions
\begin{align}
\gamma_4(2) & = -2496 + 576 \, \zeta_3 - 1440 \, \zeta_5 \, , \nonumber \\[1 mm]
\gamma_4(4) & = -\frac{8045275}{2187} + \frac{114500}{81} \zeta_3 - \frac{25000}{9} \zeta_5 \, , \\[2 mm]
\gamma_4(6) & =-\frac{393946504469}{91125000} + \frac{11736088}{5625} \zeta_3 - \frac{19208}{5} \zeta_5 \, , \nonumber \\[2 mm]
\gamma_4(8) &= -\frac{5685358151649447407}{1200725694000000} + \frac{
 142906863577}{54022500} \zeta_3 - \frac{1158242}{245} \zeta_5 \nonumber
 \end{align}
 in full agreement with the integrability prediction in the literature \cite{Kotikov:2007cy,Bajnok:2008qj}. The results for the structure constants are
 \begin{align}
 \alpha_4(2) & = 9952 + 1312 \, \zeta_3 + 288 \, \zeta_3^2 + 3920 \, \zeta_5 + 5880 \, \zeta_7 \, , \nonumber \\[1 mm]
 \alpha_4(4) & = \frac{1930033531879}{882165816} + \frac{15976465}{83349} \zeta_3 + \frac{1000}{21} \zeta_3^2 + \frac{795070}{1323} \zeta_5 + 700 \zeta_7 \, , \label{alphas} \\[2 mm]
 \alpha_4(6) & =\frac{357114900616418917}{1320819513750000} + \frac{15290724568}{1010728125} \zeta_3 + \frac{1372}{275} \zeta_3^2 + \frac{440377}{7425} \zeta_5 + \frac{686}{11} \zeta_7 \, , \nonumber \\[2 mm]
 \alpha_4(8) &=\frac{3842713470388383550340207796269}{143599850269554562884000000000} + \frac{15963568233610679701}{17914135099360500000} \zeta_3  \nonumber \\[2 mm]
 &+ \frac{1158242}{2627625} \zeta_3^2 + \frac{11715335287}{2347663500} \zeta_5 + \frac{10654}{2145} \zeta_7 \, . \nonumber
\end{align}
As in lower loop orders we observe that only odd $\zeta$ values contribute, $\pi$ is absent. We expect the fourth anomalous dimension to have transcendentality weight seven. There is in fact no $\zeta_7$ term. This is explained by the $x - \bar x$ factors that probably every single scalar conformal in our list will show in the denominator (for non-ladder four-loop examples see \cite{Drummond:2013nda,Eden:2016dir}): $\log(u) \, \zeta_7$ is of weight eight, so it cannot be multiplied by another Goncharov log. Such a term would cause a singularity at $x = \bar x$, which is not a feature of any Feynman integral. For the same reason there is no constant $\zeta_3 \,\zeta_5$ term --- or indeed any other constant of the same weight --- in the structure constants which we may expect to have polylogarithm weight eight \cite{Eden:2012rr}.

Our analysis confirms the result of \cite{Goncalves:2016vir} for $\alpha_4(2)$, which is in itself a valuable piece of information because the calculation is large enough to offer space for errors. Yet, the main incentive for this project was to provide data for the verification and extension of the hexagon proposal \cite{Basso:2015zoa} for planar correlation functions in ${\cal N}= 4$ SYM. We can certainly claim to have succeeded: if the result of \cite{Goncalves:2016vir} gave five data points to compare with (the various rational numbers in the first line of (\ref{alphas})), we now have twenty. In this vein, perhaps it is worth pushing the expansion by regions to slightly higher values of the spin. Nevertheless, we would not arrive at a closed form expression in terms of harmonic sums as in \cite{Eden:2012rr} because the weight eight ansatz contains far more constants than one may hope to fix by comparing to IBP results.

\section*{Acknowledgments.}

We thank Vladimir Smirnov for many discussions about this project and a feasibility test, and the Humboldt-University group for supramolecular systems for the permission to profit from their computing resources. BE acknowledges support by SFB 647 of the DFG, and the Cluster of Excellence ``Image, Knowledge, Gestaltung'' at Humboldt-University Berlin, funded by the Excellence Initiative and DFG.

\end{document}